\documentclass[journal=jmcmar]{achemso}

\usepackage[version=3]{mhchem}
\usepackage[T1]{fontenc} 
\usepackage{graphicx}
\usepackage{dcolumn}
\usepackage{pifont}
\usepackage{bm}
\usepackage{multirow}
\usepackage{amsmath}
\usepackage{float}
\usepackage{txfonts}

\usepackage[dvips]{color}
\definecolor{darkgreen}{rgb}{0,.6,0}

\author{Yong-Man Jang}
\affiliation{Department of Organic Chemistry, Faculty of Chemistry, Kim Il Sung University, Ryongnam-Dong, Taesong District, Pyongyang, Democratic People's Republic of Korea}

\author{Chol-Jun Yu}
\email{ryongnam14@yahoo.com}
\author{Jin-Song Kim}
\affiliation{Computational Materials Design (CMD), Faculty of Materials Science, Kim Il Sung University, Ryongnam-Dong, Taesong District, Pyongyang, Democratic People's Republic of Korea} 

\author{Song-Un Kim}
\affiliation{Department of Organic Chemistry, Faculty of Chemistry, Kim Il Sung University, Ryongnam-Dong, Taesong District, Pyongyang, Democratic People's Republic of Korea}

\title{{\it Ab initio} design of drug carriers for zoledronate guest molecule using phosphonated and sulfonated calix[4]arene and calix[4]resorcinarene host molecules}

\begin{document}

\begin{tocentry}
\includegraphics[clip=true,width=2in]{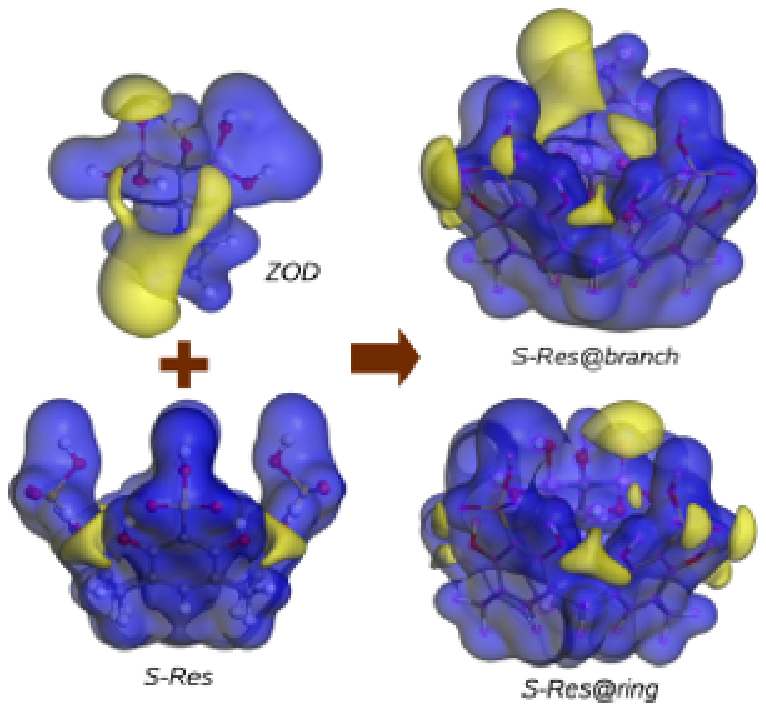} 
\end{tocentry}

\begin{abstract}
Monomolecular drug carriers based on calix[n]-arenes and -resorcinarenes containing the interior cavity can enhance the affinity and specificity of the osteoporosis inhibitor drug zoledronate (ZOD). In this work we investigate the suitability of nine different calix[4]-arenes and -resorcinarenes based macrocycles as hosts for the ZOD guest molecule by conducting {\it ab initio} density functional theory calculations for structures and energetics of eighteen different host-guest complexes. For the optimized molecular structures of the free, phosphonated, sulfonated calix[4]-arenes and -resorcinarenes, the geometric sizes of their interior cavities are measured and compared with those of the host-guest complexes in order to check the appropriateness for host-guest complex formation. Our calculations of binding energies indicate that in gaseous states some of the complexes might be unstable but in aqueous states almost all of the complexes can be formed spontaneously. Of the two different docking ways, the insertion of ZOD with the \ce{P-C-P} branch into the cavity of host is easier than that with the nitrogen containing heterocycle of ZOD. The work will open a way for developing effective drug delivering systems for the ZOD drug and promote experimentalists to synthesize them.
\end{abstract}

\section{Introduction}
\begin{figure}[!t]
\centering
\begin{tabular}{cc}
(a)\includegraphics[clip=true,scale=0.6]{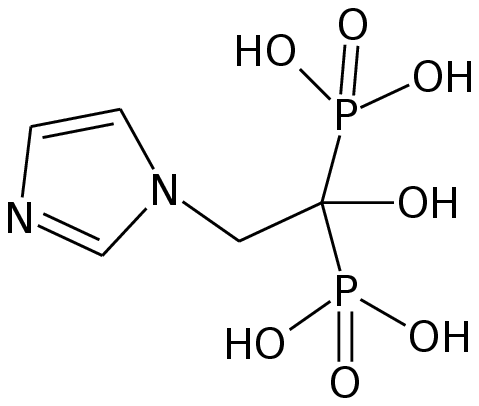} &
(b)\includegraphics[clip=true,scale=0.6]{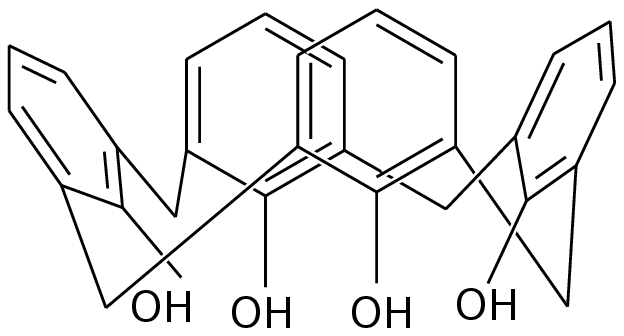} \\
\end{tabular}
\caption{\label{fig_zodstr}Chemical structures of (a) zoledronate and (b) calix[4]arene.}
\end{figure}

Zoledronic acid (or zoledronate: ZOD) is a third generation of bisphosphonates (BPs), being widely used as a successful drug in the treatment of patients with bone diseases such as osteoporosis and metabolic bone disorders. It consists of two phosphonate groups ($-$\ce{PO(OH)2}), central carbon atom between them and two side groups of hydroxyl group and heterocyclic group containing two nitrogen atoms, as shown in Figure~\ref{fig_zodstr}(a). The operation of ZOD can be explained by that the \ce{P-C-P} backbone of ZOD has a high binding affinity to bone mineral hydroxyapatite (HAP) due to a structural analog with the \ce{P-O-P} part of HAP, and is more resistant to chemical and enzymatic hydrolysis than the \ce{P-O-P} group of pyrophosphate of HAP, while the nitrogen-containing heterocyclic group of ZOD has a strong antiresoptive potency to inhibit the activity of osteoclast~\cite{Russell11,Bartl,Green,Nancollas,Ebetino,yucj16,Merino}. With a constant continuation of seeking for effective BPs that have a strong efficacy, it is also desirable to develop a drug carrier, with which the ZOD drug can reach the disordered bone without any loss, which can reduce the amount of drug dosage and eliminate possible side effects by uneven drug biodistribution.

In fact, drug delivery with a high specific affinity towards a pathological tissue is an important complement to drug discovery~\cite{Bae,Dreher}. Many drug delivering systems with clinical approval have been developed. Supramolecular polymeric micelles or encapsulation are a traditional choice for this aim, but there still remains a challenge of including a drug molecule by a single delivering host on a 1:1 stoichiometry. Since large molecular weights may originate some diffusion inconvenience, macromolecular carriers have been suggested to address the issues such as specificity and molecular weight. Moreover, monomolecular drug carriers can provide drug molecules with a higher sustaining potency without hampering too much their diffusion between cells within the pathological tissue. Supramolecular macrocycles with a cavitand such as crown-ethers, cryptands, cyclodextrins, calixarenes and calixresorcinarenes have been developed during the past decades and attempted to utilize as drug carriers. Among them, calixarenes and calixresorcinarenes are considered to be the third generation of macrocycles beyond cyclodextrins and crown-ethers.

Calix[$n$]arenes (or calix[$4$]resorcinarenes) are a well known class of macrocyclic compounds that have a bowl or cone shaped cavity (see Figure~\ref{fig_zodstr}(b) for calix[4]arene), and can be obtained in very high yield through a condensation of formaldehyde with p-substituted phenol derivatives in basic conditions~\cite{Gutsche1,Sliwa,Redshaw}. In respect of structural characteristics, they are composed of three parts; a wide upper rim, a narrow lower rim, and a central annulus (phenol- or resorcinol-derived cyclooligomers). Both rims can be easily modified with a variety of functional groups, such as hydrophilic long-chain organic polymer at the lower rim and hydrophobic inorganic group at the upper rim. In addition, a plenty of conformations are possible for the macrocyclic backbone, of which the number increases with the number of phenolic units $n$, depending on their functionalization. For instance, the most widely studied calix[4]arene has four major distinct conformers~\cite{Hong}. All these unique structural features provide a large flexibility of calixarenes, which in turn promise versatile hosts for various guest molecules with a broad range of size in host-guest chemistry. Their abilities to form host-guest complexes make it possible to use in nonlinear optical chromophores, ion receptors, ion selective electrodes and molecular recognition devices~\cite{Mutihac,Joseph,Ikeda}, but they have attracted much more interest in drug design and delivery~\cite{Rodik,Varejao,Hoskins,Morozova}. Indeed, the amphiphilic calixarenes, containing two rims as antipodal blades (hydrophobic and hydrophilic segments linked by noncovalent bond), can serve as carriers for many guest drugs employing the host-guest interaction. Moreover, water-soluble calixarenes due to the hydrophilic lower rim can include hydrophobic drug molecule from aqueous solution with their hydrophobic upper rim, and thus enhance solubility and bioavailability of the drug. When the host-guest system contacting with the targeted cell, the drug molecule interacts with the cell, while the remained calixarenes should be eliminated from the body without any side effect. In this context, it is also fortunate that calixarenes show little non-toxicity~\cite{Perret,Martin,Rodik,Shahgaldian}. Drug delivery and solubilization using calix[n]arenes or calix[4]resorcinarenes have been reported~\cite{Knoerr,Wheate,Arroyo,Adhikari15,Adhikari14,Bayrakci11,Bayrakci12,Yang,Tu,Sautrey, Huggins,Muro,Yang,Wang,Guo,Guo2}. However, the use of calixarenes and calixresorcinarenes as drug delivery is still a relatively novel concept, and the number of experimental or computational studies reported so far is relatively low, despite that these macrocyclic compounds possess several advantages such as ease of synthesis and modification and low toxicity.

In this work, we consider the free, phosphonated and sulfonated calix[4]-arenes and -resorcinarenes as hosts, the ZOD molecule as a guest and their possible host-guest complexes to get an insight for the formation of host-guest complexes. Systematic density functional theory (DFT) calculations are carried out for structural optimizations and energetics in combination with a conformation searching method, as already simulation works reported for interactions between calixarenes and other drug molecules~\cite{Murillo1,Murillo2,Makrlik1,Makrlik2,Makrlik3,Bayrakdar,Venk,Suwat,Hermann,Schubert,Carter}.

\section{Computational method}
\begin{figure}[!t]
\centering
\includegraphics[clip=true,scale=0.4]{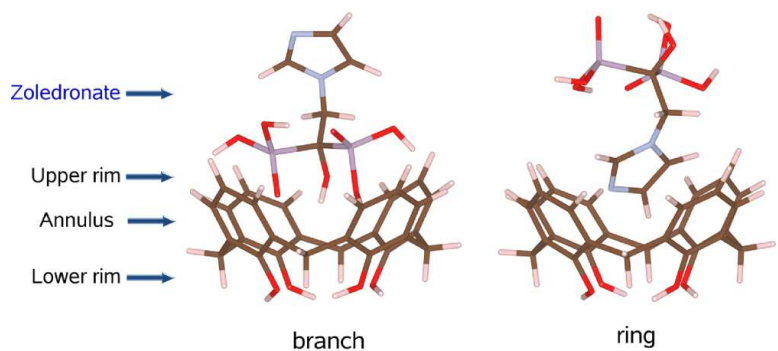}
\caption{\label{fig_docking}Docking orientations of ZOD within the host cavities of calix[4]arene.}
\end{figure}

We consider calix[4]arene and calix[4]resorcinarene as parent hosts, while applying phosphonated ($-$\ce{PO(OH)2}) and sulfonated ($-$\ce{SO2OH}) modifications at the upper rim for both cases, and methyl group ($-$\ce{CH3}) modification at the lower rim for the latter case to build further plausible hosts. Among four different conformations for calix[4]arene~\cite{Hong}, only the so-called cone conformation was utilized since this conformation has the lowest total energy compared with other conformations. Using the ZOD molecule as a guest, we build the host-guest complexes by manually docking the guest through either the \ce{P-C-P} branch or the nitrogen-containing heterocyclic ring of ZOD inside the host cavity from the upper rim, as depicted in Figure~\ref{fig_docking}. We suggest monomolecular drug carrier host and ZOD drug guest on a 1:1 stoichiometry in this work. From a geometric consideration, the insertion of the drug molecule from the lower rim can not be hypothesized. The above procedure yields nine different hosts and eighteen host-guest complexes. In order to denote the complexes concisely, we adopt a notation of {\it host@guest}, where for the host $Cal$ and $Res$ are used for calix[4]arene and calix[4]resorcinarene, prefix P- and S- for the phosphonated and sulfonated modifications, suffix -m for methyl modification, and for the guest $branch$ and $ring$ are used to distinguish the two different docking ways of ZOD. For instance, {\it S-Res-m@branch} is for the complex of sulfonated calix[4]resorcinarene combined with ZOD at the \ce{P-C-P} branch.

The resulting crude structures are first refined through a conformational search to derive the lowest energy conformation by using the Conformers module in the Materials Studio package. Stochastic searching approach that employs random generation of conformational parameters were chosen as an algorithm, where each torsion angle is randomly assigned a value in the certain range. The obtained conformations are then optimized using the {\it ab initio} pseudopotential-pseudo atomic orbital (PAO) method within the framework of DFT, as implemented in the SIESTA code~\cite{SIESTA}. We constructed the Troullier-Martins type norm-conserving pseudopotentials~\cite{TMpseudo} of all atoms using the ATOM code provided in the SIESTA package, where the valence electronic configurations are C-$2s^22p^23d^04f^0$, O-$2s^22p^43d^04f^0$, H-$1s^12p^03d^04f^0$, S-$3s^23p^43d^04f^0$, N-$2s^22p^33d^04f^0$, and P-$3s^23p^33d^04f^0$. Transferability testings are performed for these generated pseudopotentials. The standard double-$\zeta$ and polarized (DZP) basis sets for PAOs of all the atoms are used, where the energy shift for orbital-confining cutoff radii is 300 meV and the split norm for the split-valence of basis is 0.25. The BLYP~\cite{Becke,LYP} functional within the generalized gradient approximation (GGA) is used for the exchange-correlation interactions between the valence electrons. We consider dispersive van der Waals (vdW) interactions between molecules by employing the semi-empirical Grimme's approach and using the necessary parameters provided therein~\cite{Grimme}. The energy cutoff to set the wavelength of the shortest plane wave represented on the real grid, which controls the mesh size of grid, is set to be 300 Ry. In the structural optimizations, the atoms are relaxed until the exerted forces converge less than 0.02 eV/\AA. The three-dimensional periodic supercells with cubic symmetry are used to make modeling for isolated molecules. Lattice constants of these cubic supercells are 40 \AA, which can provide long distance ($\sim$30 \AA) enough to prevent artificial interactions between periodic images of molecule. 

\section{Result and discussion}
\begin{figure}[!t]
\centering
\begin{tabular}{r}
\includegraphics[clip=true,scale=0.4]{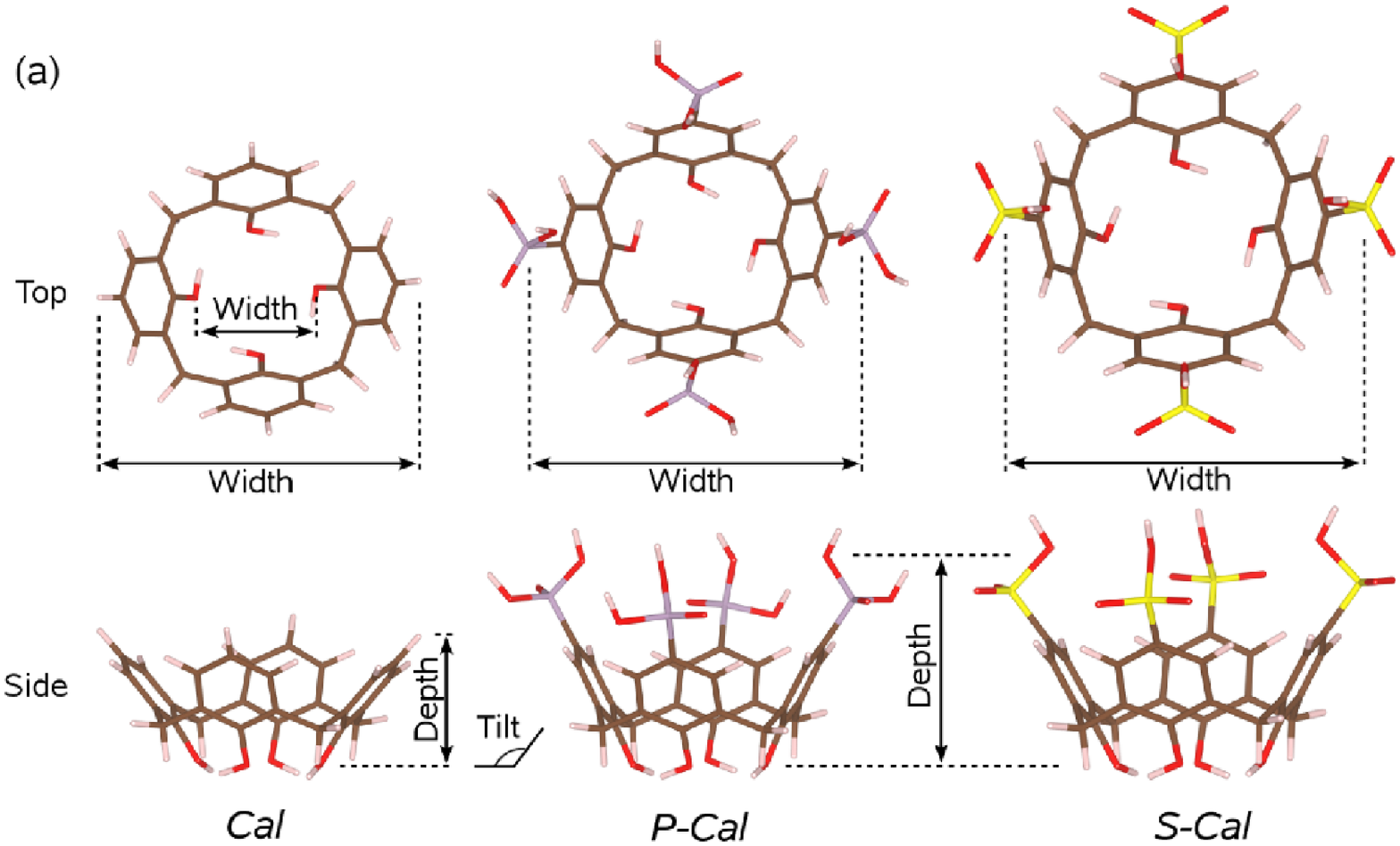} \\
\hline \\
\includegraphics[clip=true,scale=0.4]{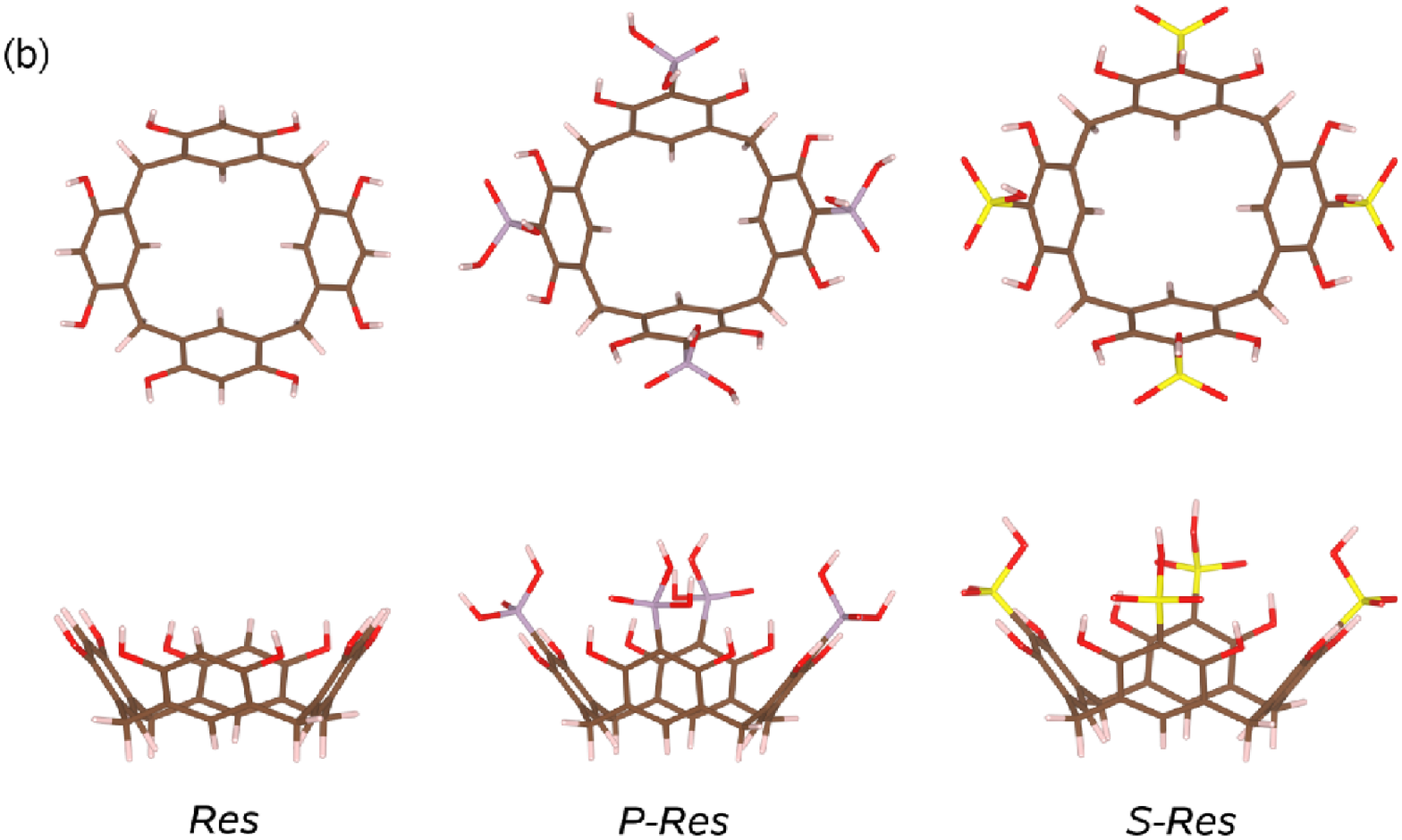} \\
\hline \\
\includegraphics[clip=true,scale=0.4]{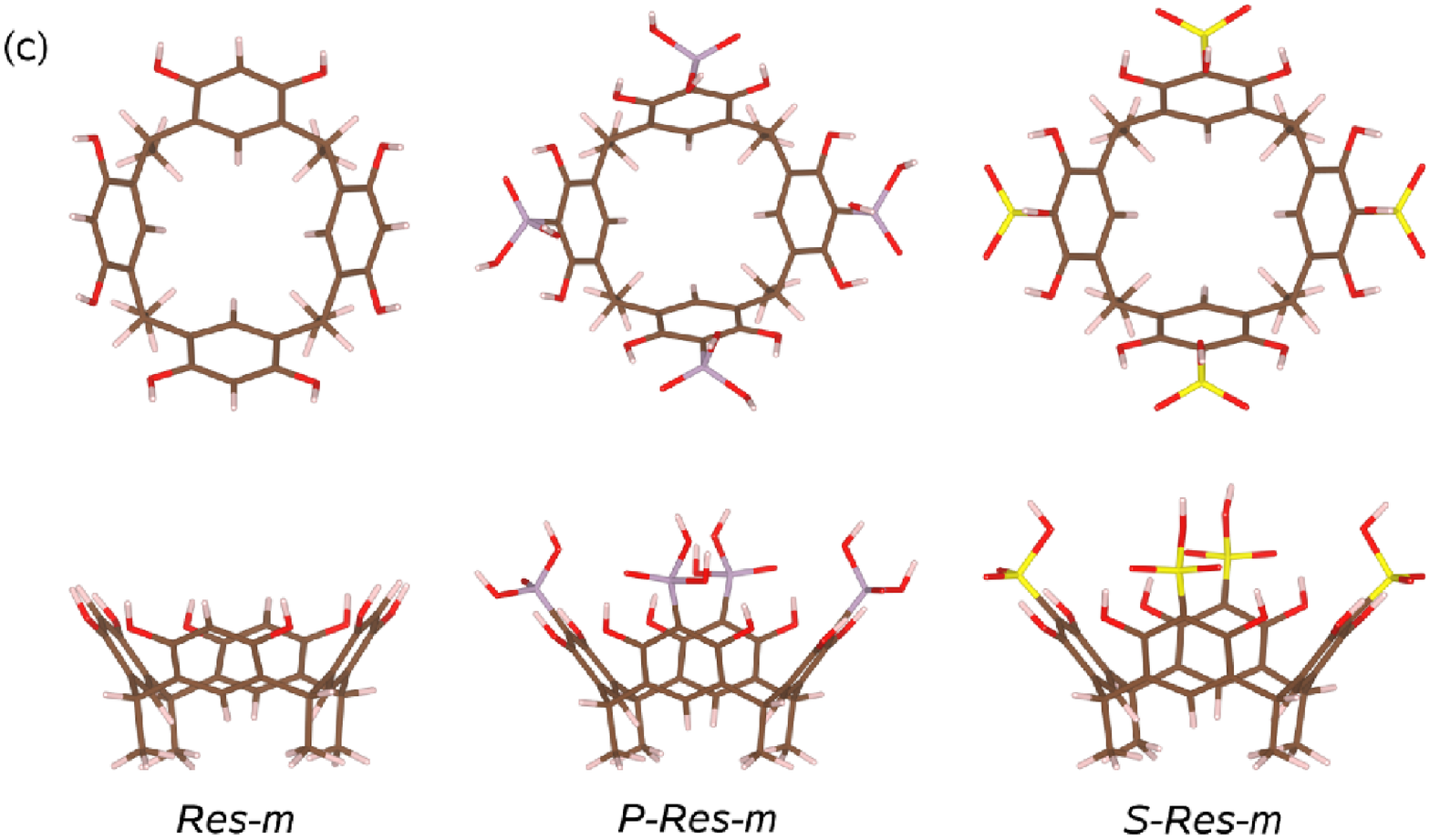} \\
\hline
\end{tabular}
\caption{\label{fig_host}Atomistic structures of the hosts optimized with DZP basis sets and BLYP + vdW (Grimme) functional. (a) Free, phosphonate and sulfonate calix[4]arenes, (b) calix[4]resorcinarenes and (c) calix[4]resorcinarenes with \ce{CH3} modification at the lower rim. Top and side views are presented.}
\end{figure}

We first present the structures of the nine host macrocycles as drug carriers, optimized with DZP basis sets and the BLYP-GGA functional with the inclusion of vdW correction. Figure~\ref{fig_host} shows the optimized atomistic structures of calix[4]arene and calix[4]resorcinarenes, which are the most familiar and widely studied calixarenes, and their phosphonate and sulfonate derivatives.
\begin{table}[!b]
\caption{\label{tab_hostgeo}Width, depth and tilt angle in the interior cavity of the 9 host compounds, depicted in Figure~\ref{fig_host}.}
\begin{tabular}{lcccc}
\hline
         & \multicolumn{2}{c}{Width (\AA)} & Depth & Tilt angle \\
\cline{2-3} 
Compound & Upper & Lower &  (\AA) & (deg) \\
\hline
Cal     & 10.06 & 4.12 & 4.35 & 124 \\
P-Cal   & 10.47 & 4.16 & 6.37 & 122 \\
S-Cal   & 10.34 & 4.20 & 6.28 & 121 \\
\hline
Res     & 10.32 & 4.22 & 3.86 & 128 \\
P-Res   & 10.51 & 4.37 & 6.05 & 123 \\
S-Res   & 10.72 & 4.27 & 5.86 & 125 \\
\hline
Res-m   & 10.36 & 4.17 & 3.81 & 129 \\
P-Res-m & 10.49 & 4.33 & 6.00 & 123 \\
S-Res-m & 10.76 & 4.27 & 5.86 & 126 \\
\hline
\end{tabular}
\end{table}
We only considered the cone conformation among four different conformations such as cone, partial cone (paco), 1,2-alternate and 1,3-alternate~\cite{Gutsche1,Hong}, considering that it is the most stable due to the lowest total energy and in particular it has the interior cavity with a shape of bowl that is necessary to form a host-guest couple. Calix[4]arenes consist of four phenolic residues bridged by methylene groups and hydroxyl groups forming hydrogen bonds at the lower rim, whereas calix[4]resorcinarenes are a related class of cyclic tetramers prepared by reacting resorcinol with an aldehyde. It is possible to fix the calix[4]resorcinarene into the cone-like structure by adding additional methylene groups between the phenolic oxygen atoms, forming a generic structure known as a cavitand. The interior cavity of calix[4]arene and calix[4]resorcinarene, which is in general hydrophobic in nature, should be sufficiently deep or wide in order to embrace small organic drug molecule, forming reversible host-guest complexes, and this can be achieved by adding functional groups such as phosphonate or sulfonate at the upper rim as we did in this work. We measure the geometrical characteristics (width at the upper and lower rims, depth, and tilt angle, as depicted in Figure~\ref{fig_host}(a)) of the interior cavity. Table~\ref{tab_hostgeo} shows the corresponding average values in the optimized structures. It was observed that the largest upper width are found in {\it P-Cal} (10.47 \AA), {\it S-Res} (10.72 \AA) and {\it S-Res-m} (10.76 \AA), while the largest lower width is found in {\it S-Cal} (4.20 \AA), {\it P-Res} (4.27 \AA) and {\it P-Res-m} (4.33 \AA). The phosphonate and sulfonate derivatives have $\sim$2 \AA~larger depth than free compounds, and the phosphonate derivatives have slighter larger depth than the sulfonate ones. In addition, the cavity of free hosts is a little more inclined than that of functionalized hosts. When compared with the geometry of ZOD molecule that has longer width of 7.13 \AA~at the branch, width of 4.22 \AA~at the ring, and length of 7.11 \AA, all the nine hosts have the appropriate cavity size.

Then we analyze the atomistic structures of the host-guest complexes comprised of the above mentioned hosts and the guest ZOD molecule. There are two different docking ways according to the orientation of ZOD insertion inside the cavity in each case of the 9 different hosts, (1) insertion of ZOD with the \ce{P-C-P} branch pointing inwards the cavity in the direction of the lower rim and (2) with the heterocyclic ring in the same way as for the first case, resulting in the different 18 host-guest complexes. In Table~\ref{tab_hosgus_cal} we show the geometric features of the interior cavity in the supramolecular host-guest complexes. It was found that when compared with two different docking ways for the cases of free hosts, there is little change in the upper width (and the tilt angle) of the cavity, which might be the most changeable geometric factor upon the host-guest coupling, while the lower width and depth in the $branch$ docking way are smaller and larger than in the $ring$ way for the resorcinarenes although little change for the calixarenes. Meanwhile, the cavity of the phosphonato- and sulfonato-hosts coupled with the ZOD guest in the $branch$ way is clearly wider at the upper rim, shallower, and more inclined than in the $ring$ docking way, whereas there is no distinction in lower widths between the two different docking ways. On the other hand, when compared with the hosts before coupling, it can be said that there is little change in the cavity geometry for the free hosts upon host-guest formation, although small amount of changes were observed in the lower width. On the contrary, for cases of the hosts functionalized with phosphonate and sulfonate at the upper rim, the cavity became significantly wider, deeper and more inclined upon insertion of ZOD molecule. These indicate that there is strong interaction and thus binding between the functionalized hosts and ZOD molecule, while weak interaction for the free hosts.
\begin{table}[!t]
\caption{\label{tab_hosgus_cal}Width, depth and tilt angle in the interior cavity of the 18 host-guest complexes.}
\begin{tabular}{lccccc}
\hline
         & \multicolumn{2}{c}{Width (\AA)} & Depth & Tilt angle \\
\cline{2-3} 
Compound & Upper & Lower &  (\AA) & (deg) \\
\hline
Cal@branch   & 10.06 & 4.16 & 4.38 & 124 \\
Cal@ring     & 10.06 & 4.13 & 4.35 & 124 \\
P-Cal@branch & 10.92 & 3.99 & 6.25 & 126 \\
P-Cal@ring   & 10.79 & 4.07 & 6.31 & 124 \\
S-Cal@branch & 10.93 & 4.09 & 6.14 & 125 \\
S-Cal@ring   & 10.82 & 4.11 & 6.24 & 125 \\
\hline
Res@branch   & 10.32 & 4.20 & 3.86 & 128 \\
Res@ring     & 10.32 & 4.53 & 3.96 & 126 \\
P-Res@branch & 10.76 & 4.56 & 5.96 & 124 \\
P-Res@ring   & 10.63 & 4.50 & 6.06 & 123 \\
S-Res@branch & 10.91 & 4.27 & 5.74 & 127 \\
S-Res@ring   & 10.79 & 4.58 & 5.95 & 124 \\
\hline
Res-m@branch   & 10.37 & 4.35 & 3.86 & 128 \\
Res-m@ring     & 10.38 & 4.57 & 3.92 & 127 \\
P-Res-m@branch & 10.81 & 4.53 & 5.93 & 124 \\
P-Res-m@ring   & 10.63 & 4.50 & 6.06 & 123 \\
S-Res-m@branch & 10.91 & 4.41 & 5.81 & 126 \\
S-Res-m@ring   & 10.85 & 4.74 & 5.97 & 124 \\
\hline
\end{tabular}
\end{table}
\begin{figure}[!t]
\centering
\begin{tabular}{c}
\includegraphics[clip=true,scale=0.4]{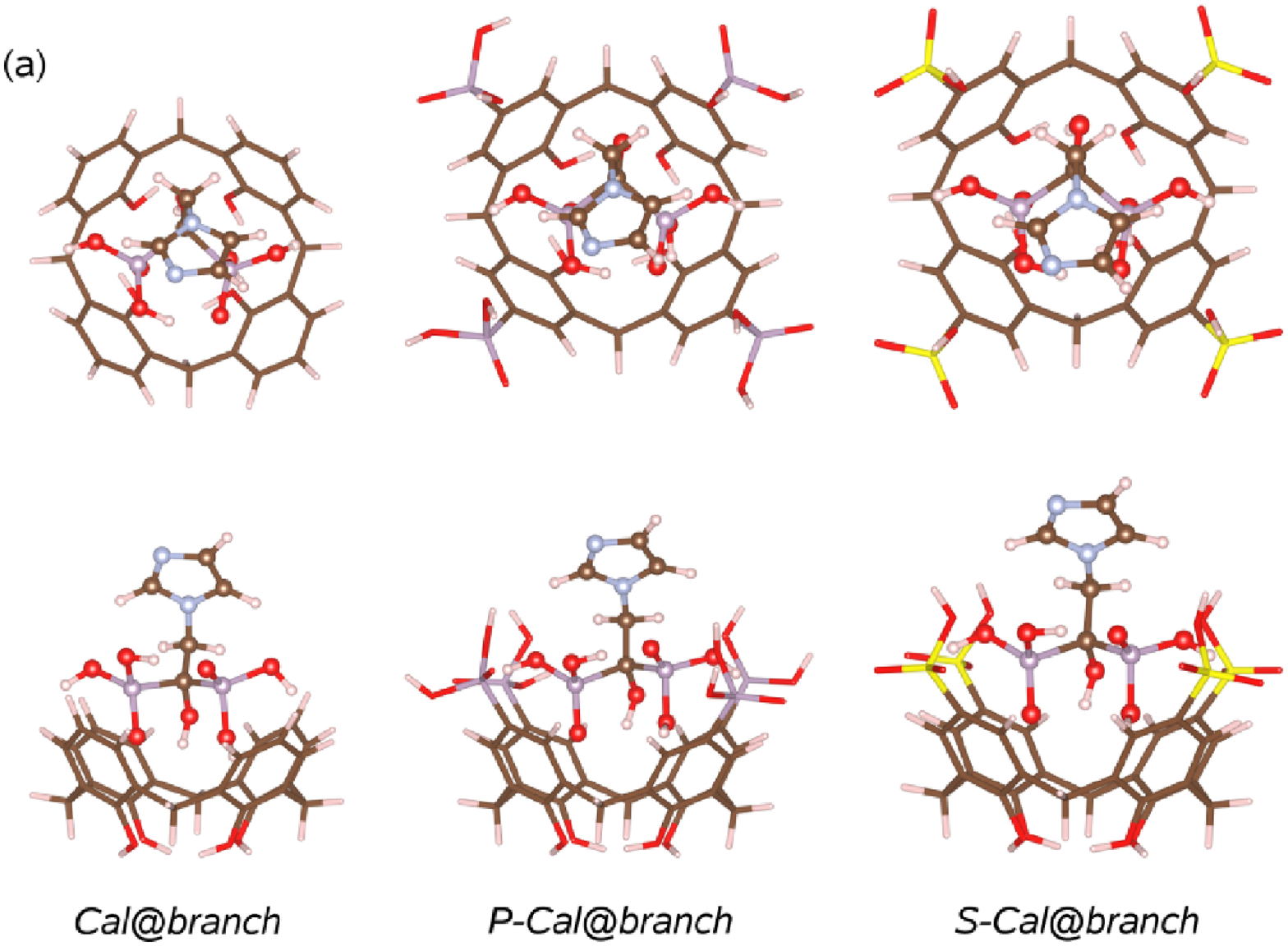}\\
\hline \\
\includegraphics[clip=true,scale=0.4]{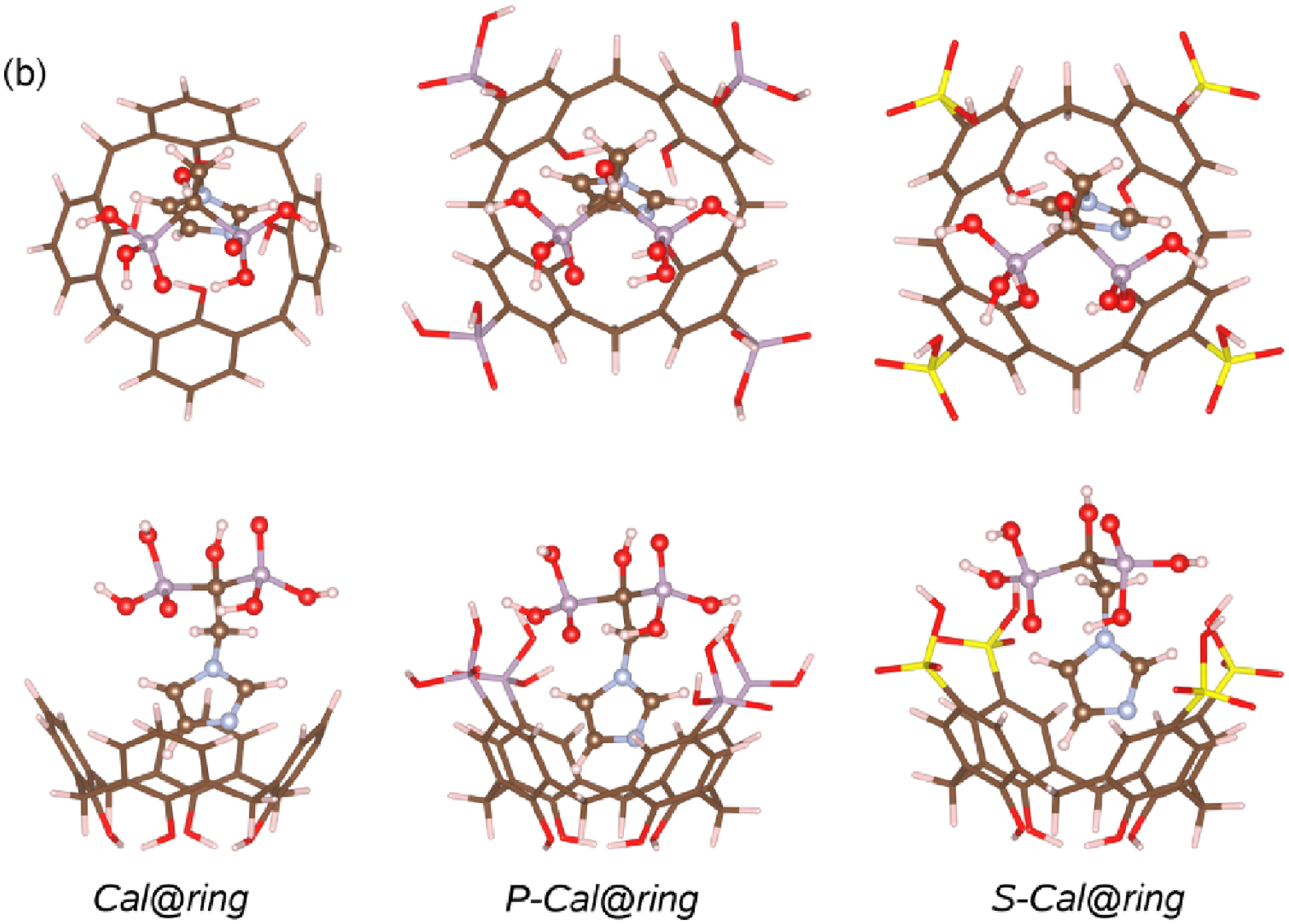}\\
\hline
\end{tabular}
\caption{\label{fig_cal-zod}Optimized structures of host-guest complexes formed by free, phosphonato- and sulfonato-calix[4]arenes as hosts and ZOD as guest in top (upper panel) and side (lower panel) views. Docking is constructed by (a) the \ce{P-C-P} branch of ZOD and (b) the heterocyclic ring coupled with the upper rim of calixarenes.}
\end{figure}
Figure~\ref{fig_cal-zod} shows the optimized structures of host-guest complexes formed by free, phosphonato- and sulfonato-calix[4]arenes and ZOD molecule. For the cases of $branch$ docking ways, we can see several hydrogen bonds between ZOD branch oxygen atoms and hydrogen atoms of the host upper rim, and additionally for the cases of phosphonate and sulfonate derivatives between ZOD hydrogen atoms and oxygen atoms of the host upper rim, resulting in the hydrogen-mediated binding between host and guest. We measured hydrogen bond lengths for the former hydrogen bond as 2.69, 2.86 \AA~for {\it Cal@branch}, 2.84, 3.10 \AA~for {\it P-Cal@branch}, and 2.90, 3.04, 3.10 \AA~for {\it S-Cal@branch}, while for the latter hydrogen bond as 2.19, 2.60, 2.88 \AA~for {\it P-Cal@branch} and 2.54, 2.46 \AA~for {\it S-Cal@branch}. For the case of $ring$ docking way, meanwhile, the hydrogen bond was formed between the ZOD branch and the host upper rim for functionalized calix[4]arenes, whereas it can not be formed in {\it Cal@ring} due to quite long separation ($>$3.5 \AA) between the ZOD ring nitrogen and the host upper rim hydrogen. The hydrogen bond lengths were measured to be 2.08, 2.76, 2.92 \AA~for {\it P-Cal@ring} and 2.57, 2.59, 2.60, 2.64, 2.65, 3.01 \AA~for {\it S-Cal@ring}, which indicates that the latter case has stronger binding than the former case due to more hydrogen bonds. It is worth noting that the distance between the ZOD ring carbon atoms and the host phenyl carbon atoms are 2.97, 3.20, 3.50 \AA~for {\it Cal@ring}, 3.09, 3.26, 3,34 \AA~for {\it P-Cal@ring} and 3.47, 3.50 \AA~for {\it S-Cal@ring}, indicating $\pi-\pi$ stacking interaction between the ZOD heterocyclic ring carbon atoms and the host phenyl ring carbon atoms. Therefore, we can predict that the host-guest formation with the ring docking way is preferable to the branch docking way with respect to the intermolecular interaction.

\begin{figure}[!t]
\centering
\begin{tabular}{c}
\includegraphics[clip=true,scale=0.3]{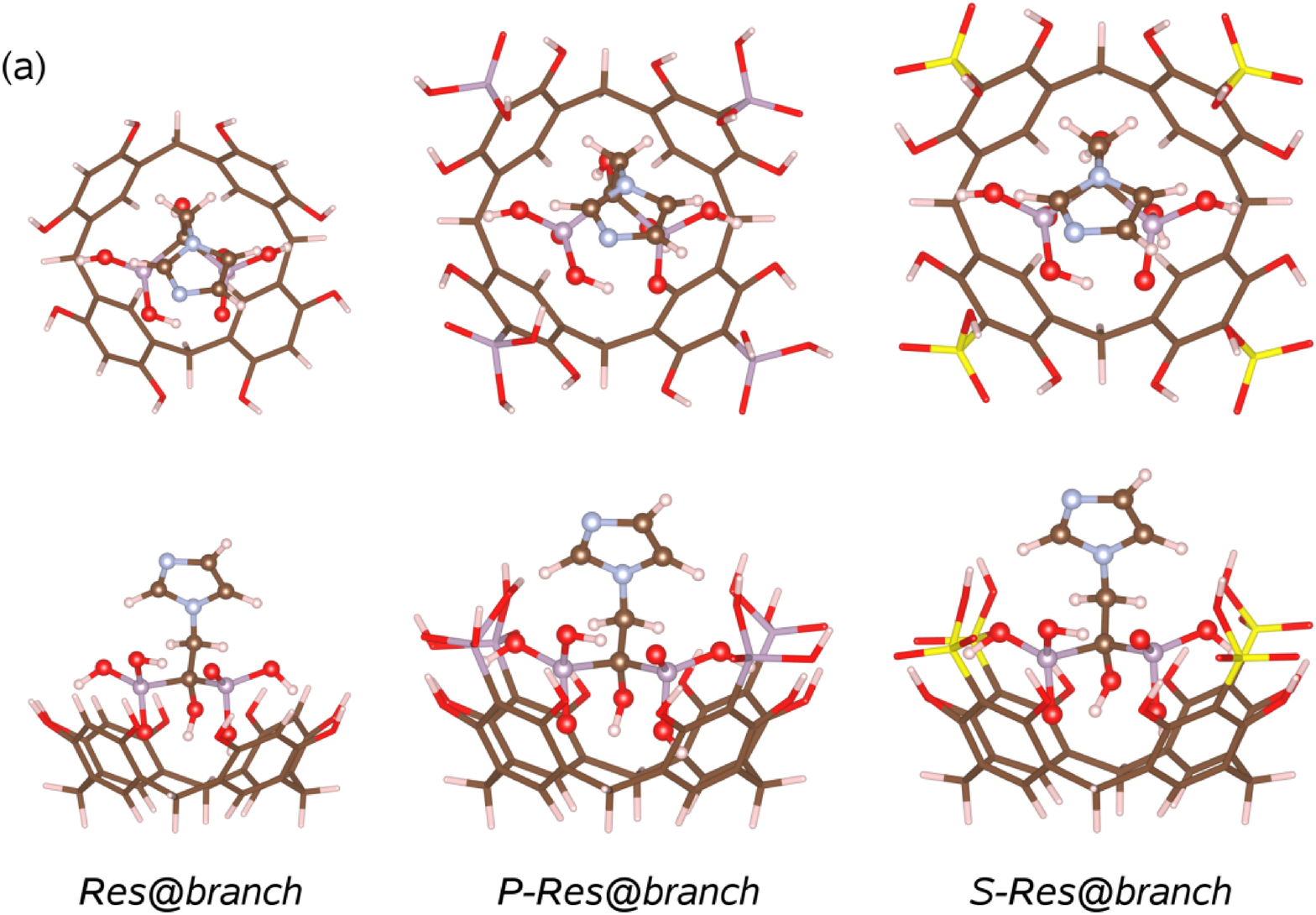}\\
\hline \\
\includegraphics[clip=true,scale=0.3]{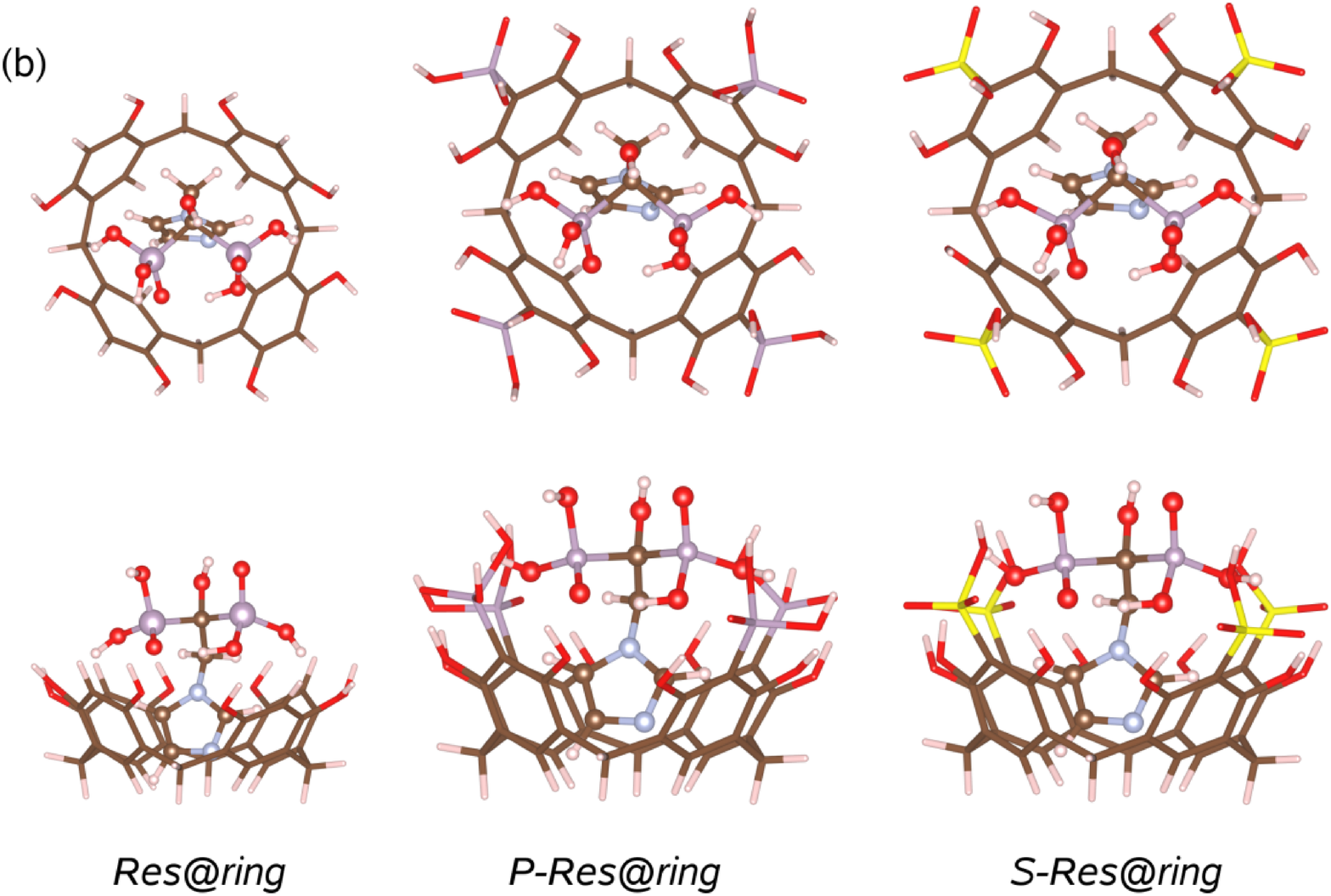}\\
\hline
\end{tabular}
\caption{\label{fig_res-zod}Optimized structures of host-guest complexes formed by free, phosphonato-and sulfonato-calix[4]resorcinarenes and ZOD.}
\end{figure}
\begin{figure}[!t]
\centering
\begin{tabular}{c}
\includegraphics[clip=true,scale=0.3]{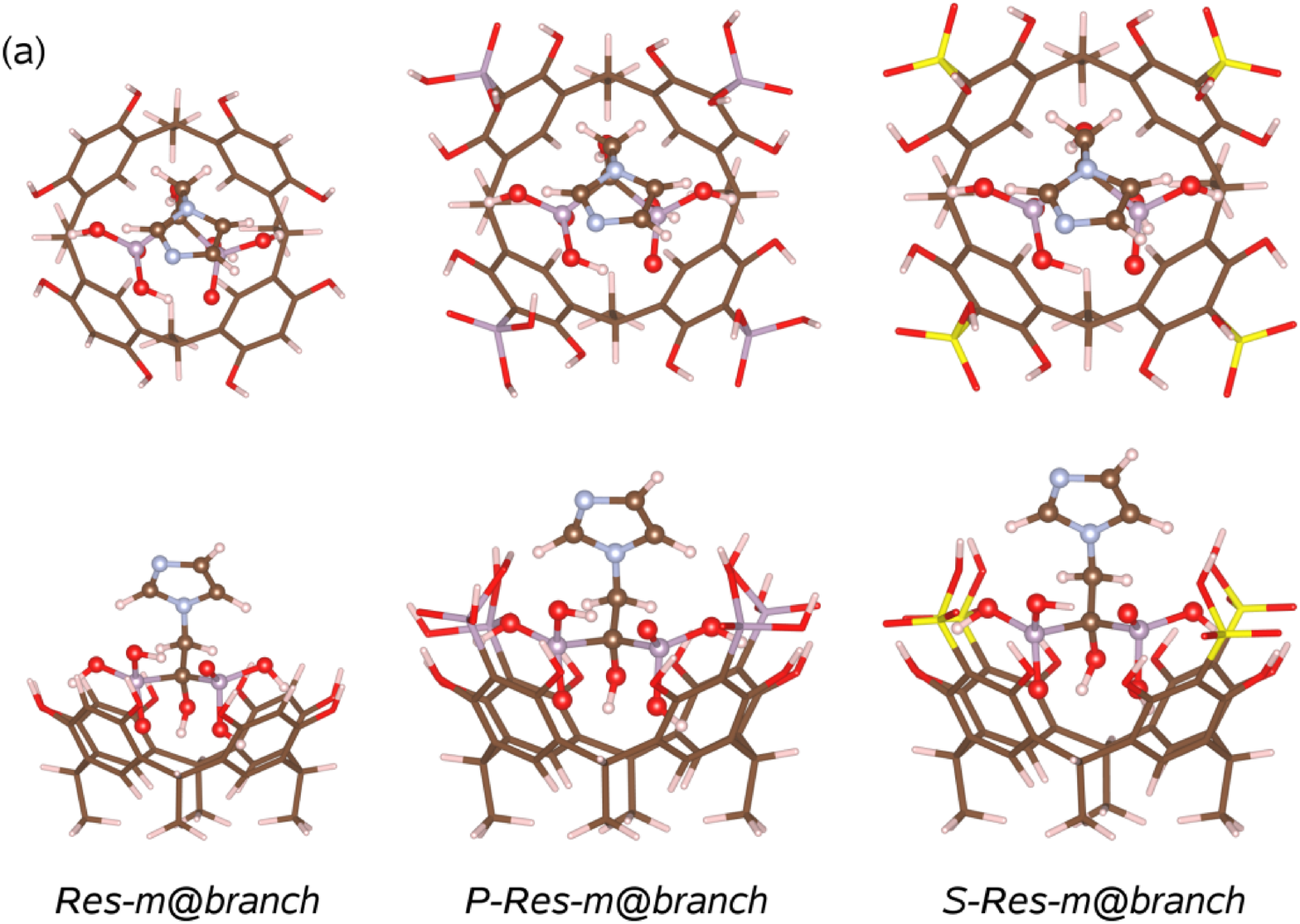}\\
\hline \\
\includegraphics[clip=true,scale=0.3]{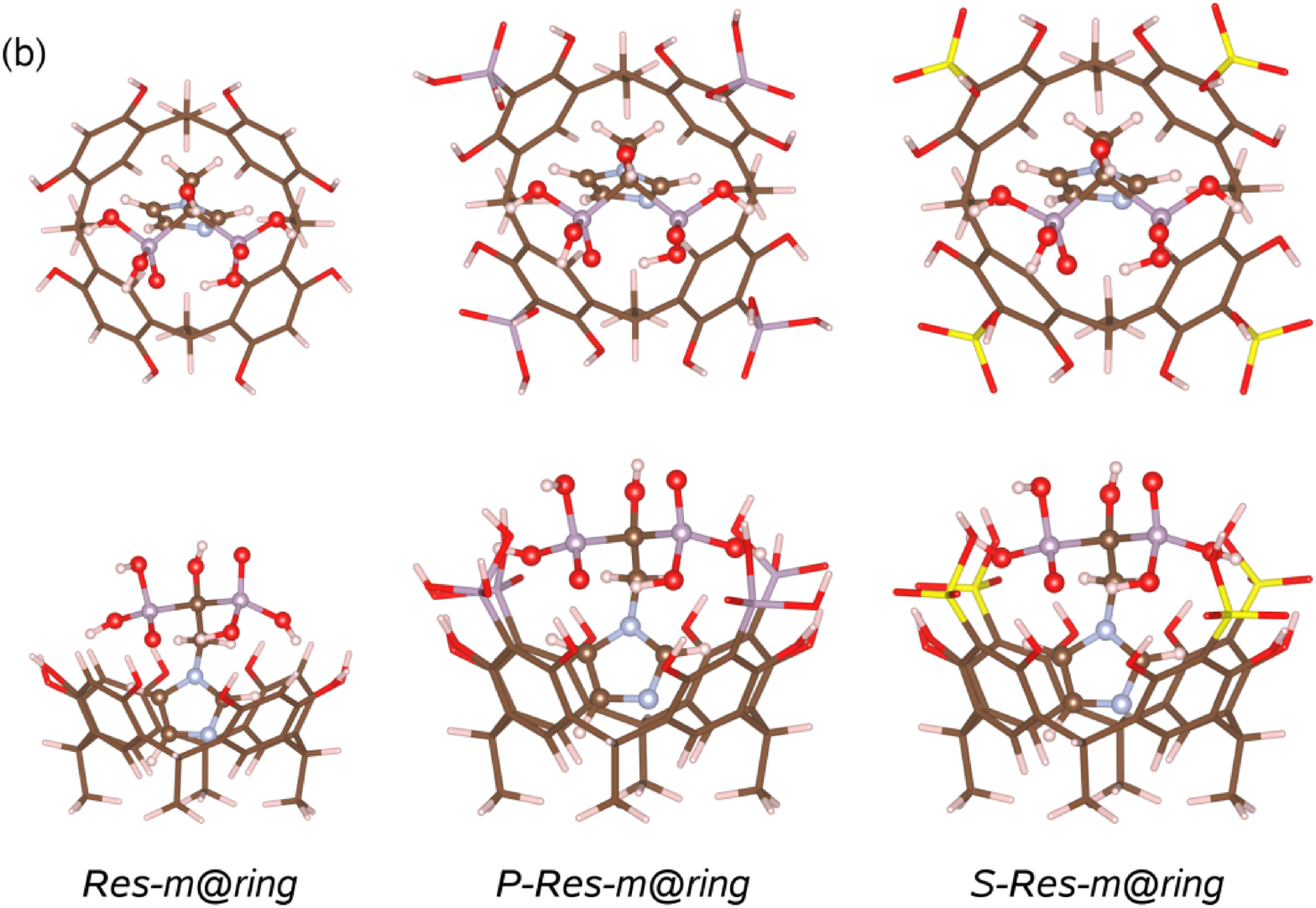}\\
\hline
\end{tabular}
\caption{\label{fig_resch3-zod}Optimized structures of host-guest complexes formed by free, phosphonato- and sulfonato-calix[4]resorcinarenes modified by \ce{CH3} group at the lower rim and ZOD.}
\end{figure}
Similar arguments to the above cases of calix[4]arenes hold for calix[4]resorcinarenes and their modifications with methyl groups $-$\ce{CH3} at the lower rim. Unlike the calix[4]arenes, they have additional OH groups at the upper rim, which are from resorcinol, and thus further hydrogen bond was formed between these O atoms and the ZOD branch H atoms. In Figure~\ref{fig_res-zod}, we show the optimized structures of host-guest supramolecular complexes comprised of free, phosphonato- and sulfonato-calix[4]resorcinarenes and ZOD molecule. The shortest \ce{O$\cdots$H} hydrogen bond lengths are 1.78, 1.95, 2.09, 2.11, 2.08, 2.07 \AA~for {\it Res@branch}, {\it P-Res@branch}, {\it S-Res@branch}, {\it Res@ring}, {\it P-Res@ring}, and {\it S-Res@ring}, respectively. Typically, these bond lengths are shorter than those in the corresponding calix[4]arene host-guest complexes, indicating stronger binding between host and guest. For the $\pi-\pi$ stacking interaction in the ring docking cases, the shortest \ce{C-C} distances are measured to be about 2.6, 2.6, 2.7 \AA~for {\it Res, P-Res} and {\it S-Res}, respectively. Figure~\ref{fig_resch3-zod} shows the optimized molecular structures of host-guest complexes from calix[4]resorcinarenes modified with methyl groups at the lower rim. Again, the hydrogen bond lengths and \ce{C-C} distances for $\pi-\pi$ stacking interaction similar to the cases of calix[4]resorcinarenes were observed. It should be noted that when compared with the {\it Cal@ring}, both {\it Res@ring} and {\it Res-m@ring} have deeper insertion of the guest ZOD molecule inside the cavity due to possibly OH groups at the upper rim, indicating stronger binding between host and guest.

\begin{figure*}[!t]
\centering
\includegraphics[clip=true,scale=0.6]{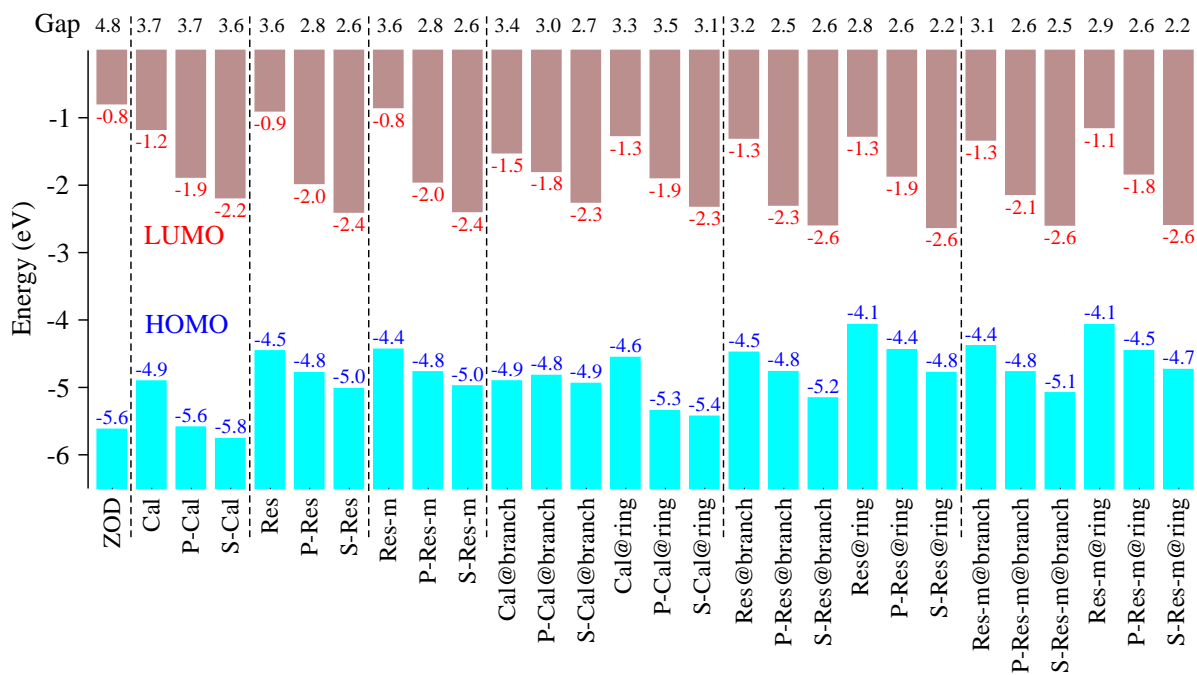}
\caption{\label{fig_fmoene}Bar chart of HOMO and LUMO energy levels in ZOD and calix[4]-arene and -resorcinarene related compounds, calculated with BLYP + vdW (Grimme) approach in aqueous state. The corresponding HOMO-LUMO energy gaps in eV unit are presented in the top.}
\end{figure*}

To sum up the structural analysis, (1) the geometrical size of the cavity of phosphonate and sulfonate modified hosts are clearly changed upon the insertion of guest ZOD molecule inside their cavity, but little change for the free hosts, (2) the calix[4]resorcinarenes support stronger binding of host-guest coupling than the calix[4]arenes due to additional hydrogen bonding by OH groups from resorcinol at the upper rim, and (3) the host-guest formations at the ring docking way is more desirable than the branch way due to additional $\pi-\pi$ stacking interaction between the ZOD ring carbon atoms and the host phenyl ring carbon atoms. However, these conclusions are only qualitative but not quantitative. The quantitative assess for the ease of host-guest formation can be done by estimating energetics such as binding energy and energy gap between the highest occupied and lowest unoccupied molecular orbital (HOMO-LUMO) in the following arguments.

The binding energy of ZOD guest to the calix[4]arenes and calix[4]resorcinarenes and their different derivatives were calculated as the difference between total energy of host-guest complexes and the total energy sum of individual host and guest molecules. In fact, it is regarded that accurate {\it ab initio} calculation of binding energies in supramolecular complexes, in which the non-covalent $\pi-\pi$ interactions between conjugated aromatic rings play a key role, is challenging due to the delicate balance between different types of intermolecular interactions~\cite{Hermann}. In this work we considered dispersive vdW interactions through empirical Grimme's approach, but we should note that more accurate description of dispersive interaction is possible with the many-body dispersion method to capture the anisotropy and collective nature of long-range correlation in supramolecular systems. Therefore, we will see only whether the host-guest complexes can be formed or not and which complex or docking way is more appropriate for the complex formation. In addition, considering that the reactions are occurred in an aqueous solution, we treated the effect of the solution on the binding energy using the conductor-like screening model (COSMO)~\cite{cosmo} with a dielectric constant 78.54 of water. Namely, we considered binding energetics in gaseous and aqueous states.

\begin{table}[!t]
\caption{\label{tab_ene}Binding energy ($E_b$) in gaseous and aqueous state, and HOMO-LUMO energy gap in the 18 host-guest complexes.}
\begin{tabular}{lrrc}
\hline
         & \multicolumn{2}{c}{$E_b$ (kcal/mol)} & Energy gap \\
\cline{2-3} 
Compound & Gaseous & Aqueous    & (eV) \\
\hline
Cal@branch   & 3.38     & $-$40.46    & 3.38 \\
Cal@ring     & 1.72     & $-$26.42    & 3.29 \\
P-Cal@branch & $-$46.86 & $-$89.34    & 3.03 \\
P-Cal@ring   & $-$39.48 & $-$85.25    & 3.46 \\
S-Cal@branch & $-$50.55 & $-$85.37    & 2.69 \\
S-Cal@ring   & $-$38.60 & $-$82.55    & 3.12 \\
\hline
Res@branch   & $-$17.94 & $-$62.82   & 3.18 \\
Res@ring     & 5.84     & $-$58.15   & 2.80 \\
P-Res@branch & 2.86     & $-$62.40   & 2.47 \\
P-Res@ring   & 1.29     & $-$62.13   & 2.58 \\
S-Res@branch & $-$34.74 & $-$92.89   & 2.57 \\
S-Res@ring   & 9.92     & $-$54.71   & 2.15 \\
\hline
Res-m@branch   & 9.67     & $-$52.20 & 3.05 \\
Res-m@ring     & 9.46     & $-$56.53 & 2.93 \\
P-Res-m@branch & 8.57     & $-$56.41 & 2.63 \\
P-Res-m@ring   & 6.30     & $-$59.80 & 2.62 \\
S-Res-m@branch & $-$32.00 & $-$91.83 & 2.49 \\
S-Res-m@ring   & 14.45    & $-$50.83 & 2.16 \\
\hline
\end{tabular}
\end{table}

The calculated binding energies in the 18 different host-guest complexes, together with their HOMO-LUMO gaps, are listed in Table~\ref{tab_ene}. We were aware of that in the gaseous state only some host-guest coupling complexes are exothermically formed, such as {\it P-Cal} and {\it S-Cal} at both branch and ring sides, {\it Res@branch}, {\it S-Res@branch}, and {\it S-Res-m@branch}, due to their negative binding energies, and the remaining compounds have positive binding energies, indicating endothermic formations. Interestingly, for the cases of calix[4]arenes, free hosts can not bind ZOD molecule, while phosphonato- and sulfonato-calix[4]arenes can bind ZOD in thermodynamically favorable way, with binding energies of $-38\sim-50$ kcal/mol. When included the aqueous solution effect, we have all of the negative binding energies. In both gaseous and aqueous states, the \ce{P-C-P} branch docking way is easier to realize host-guest formation than the heterocyclic ring way, with $2\sim3$ kcal/mol higher interaction energies for {\it P-Cal, S-Cal, P-Res, Res-m, P-Res-m}, and even $20\sim40$ kcal/mol higher interaction energies for {\it Cal, Res, S-Res} and {\it S-Res-m}. Notably, {\it S-Res@branch} and {\it S-Res-m@branch} have the largest binding energies of over $-90$ kcal/mol in aqueous state among entire host-guest complexes and even in gaseous state they have specially negative binding energies of about $-30$ kcal/mol among calix[4]resorcinarene related host-guest complexes. When compared with other calix[$n$]arene based drug carriers, calix[$n$]arene derivatives (R = $-$OEt, \ce{SO2OH}; $n$ = 4, 5, 6) coupled with a chemotherapeutic agent GTP have binding energies in the range of $-25\sim -102$ kcal/mol, and moreover, sulfonate calixarenes have superior binding energetics to ethoxyl calixarenes, {\it e.g.}, $-90.8$ kcal/mol for calix[4]arene tetrasulfonic acid with GTP~\cite{Murillo1}, in accordance well with our calculations.

\begin{figure}[!t]
\centering
\includegraphics[clip=true,scale=0.7]{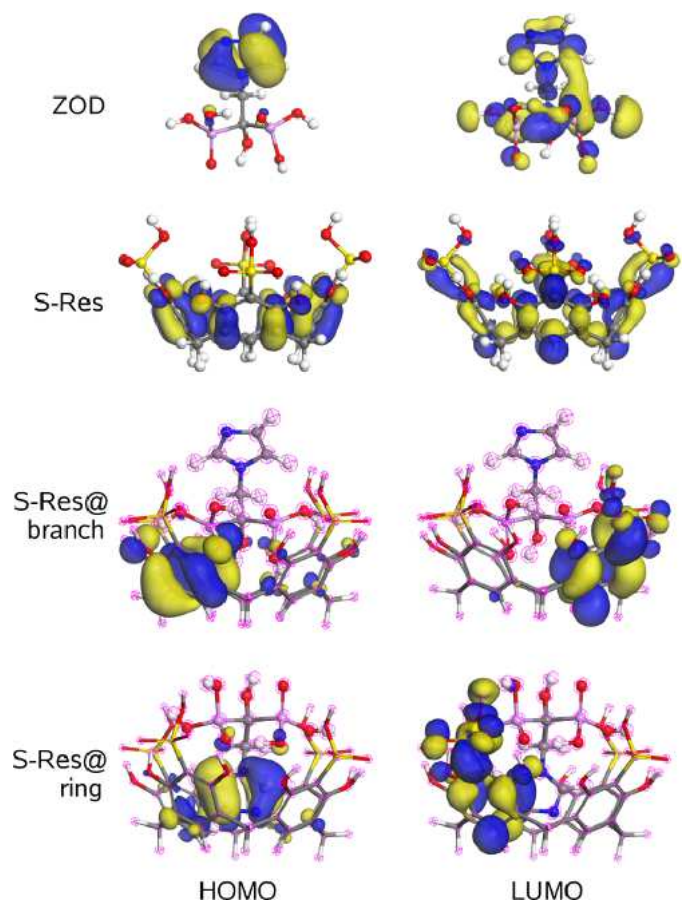}
\caption{\label{fig_ao}Isosurface map of HOMO and LUMO of ZOD molecule, sulfonato-calix[4]resorcinarene host molecule, and sulfonato-calix[4]resorcinarene coupled with ZOD at the branch and ring, evaluated at $\pm$0.03 eV. Yellow color represents the positive value, while blue color the negative value.}
\end{figure}

The frontier molecular orbitals including HOMO and LUMO are important properties that can be obtained from computational quantum chemistry. The HOMO-LUMO energy gap determines stability, chemical bonding, electrical and optical properties of the molecule. The smaller energy gap indicates in general the lower kinetic stability but the better electric conduction. It is also possible to assess the electron giving and accepting ability of the molecule by considering spatial distribution of HOMO and LUMO, which characterizes chemical bonding in the molecule. In Figure~\ref{fig_fmoene} we show the bar chart of HOMO and LUMO energy levels in all of the molecules treated in this work, together with the corresponding HOMO-LUMO energy gaps, calculated in aqueous states. At the first seeing, the order of HOMO and LUMO energy levels are free $<$ phosphonate $<$ sulfonate, but the energy gaps have the reverse order like free $>$ phosphonate $>$ sulfonate. It is found that while ZOD molecule has distinctly large energy gap of 4.8 eV (HOMO; $-$5.6 eV, LUMO; $-$0.8 eV), calix[4]-arene and -resorcinarene related hosts and host-guest complexes have smaller energy gaps in a range of $2.2\sim 3.7$ eV. And the energy gaps of the hosts are in general larger than those of the corresponding host-guest complexes. Moreover, calix[4]arenes have slightly larger energy gaps than calix[4]resorcinarenes, whereas there is no clear change between calix[4]resorcinarenes and methyl modified calix[4]resorcinarenes. It should be stressed that {\it S-Res@ring} and {\it S-Res-m@ring} have the lowest energy gaps of $\sim2.2$ eV (HOMO; $-$4.8 eV, LUMO; $-$2.6 eV). Together with the above analysis for binding energies as well as geometrical size of cavity, these let us pick the sulfonato-calix[4]resorcinarenes as the most appropriate hosts for ZOD guest molecule to make formation of {\it S-Res@branch} or {\it S-Res@ring} complexes.

\begin{figure}[!t]
\centering
\includegraphics[clip=true,scale=0.9]{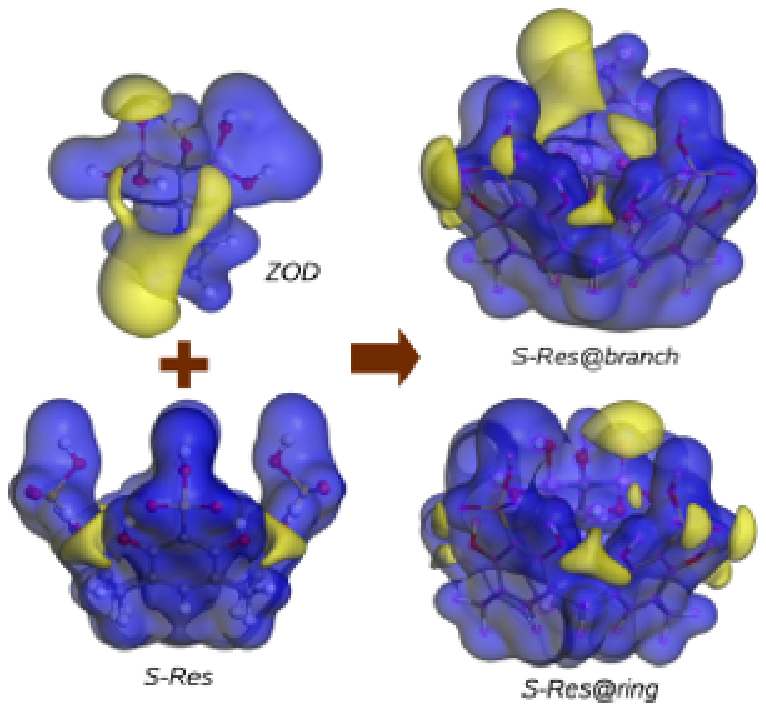}
\caption{\label{fig_pot}Isosurface map of electrostatic potential of ZOD molecule, sulfonato-calix[4]resorcinarene host molecule, and sulfonato-calix[4]resorcinarene coupled with ZOD, evaluated at 0.045 eV. Yellow color represents the positive value, while blue color the negative value. }
\end{figure}

Figure~\ref{fig_ao} shows 3-dimensional isosurface map of HOMOs and LUMOs of ZOD, {\it S-Res}, {\it S-Res@branch} and {\it S-Res@ring}. For the cases of ZOD and {\it S-Res}, HOMOs are found on the ZOD heterocyclic ring and the {\it S-Res} phenyl rings, while the majority of LUMOs on the ZOD branch and the {\it S-Res} sulfonate groups. After formation of the host-guest complexes, we observe clearly different features between branch and ring docking ways; for the branch way HOMOS and LUMOS are distributed on the opposite sides of the host annulus, while for the ring way HOMOs on the heterocyclic ring as in the case of free ZOD molecule and LUMOs on the host side. Similar features are also observed in other host-guest complexes. This indicates that for the cases of ring docking way the heterocyclic ring of ZOD plays a role of electron donor, while the phenyl ring part of hosts the electron acceptor, indicating stronger coupling than for the cases of branch docking way in respect of host-guest interaction.

The interaction between host and guest in supramolecular complexes can also be characterized by electrostatic potential map, as shown in Figure~\ref{fig_pot} for sulfonato-calix[4]resorcinarene hosts and host-guest complexes. Here, the electrophilic sites are represented as yellow color (positive value), while the nucleophilic sites are represented as blue color (negative value). The electrostatic potential map demonstrates that the most applicable atomic sites for electrophilic attack are the sulfonic group \ce{SO2OH} at the upper rim of the hosts, while the most possible sites for nucleophilic process are O$-$H groups.

\section{Conclusion}
In this work we have studied the calix[4]arenes, calix[4]resorcinarenes and their phosphonate and sulfonate derivatives as potential drug carriers for the osteoporosis inhibitor drug zoledronate in host-guest chemistry using {\it ab initio} density functional theory calculations. Norm-conserving pseudopotential pseudo atomic orbital method was adopted with the standard DZP basis sets and BLYP + vdW (Grimme) GGA functional. After refinement of structures through stochastic conformation searching, we have conducted full atomic relaxations to obtain the optimized molecular structures of selected 9 different hosts and 18 different host-guest complexes. Through careful analysis of the optimized structures, we have concluded that (1) the geometrical size of the cavity of phosphonate and sulfonate modified hosts are clearly changed upon the insertion of guest ZOD molecule inside their cavity, but little change for the free hosts, (2) the calix[4]resorcinarenes support stronger binding of host-guest coupling than the calix[4]arenes due to additional hydrogen bonding by OH groups from resorcinol at the upper rim, and (3) the host-guest formations at the ring docking way is more desirable than the branch way due to additional $\pi-\pi$ stacking interaction between the ZOD ring carbon atoms and the host phenyl ring carbon atoms. To get more quantitative assess for the host-guest complex formations, we have calculated the binding energy in gaseous and aqueous state, and the frontier molecular orbitals including HOMO and LUMO. Our calculations indicates that while in gaseous state some complexes can be formed endothermically due to positive binding energies, in aqueous state all the complexes can be formed exothermically, with larger binding energies in the branch docking way ($-52\sim -93$ kcal/mol) than the ring docking way ($-26\sim -85$ kcal/mol). The lowest HOMO-LUMO energy gap was found in sulfonato-calix[4]resorcinarenes (host) and ZOD (guest) complex at the ring docking way. Putting all accounts together, we chose the sulfonato-calix[4]arenes as the most appropriate host for ZOD guest on 1:1 stoichiometry, although other hosts are not ruled out. We believe this work will facilitate experimentalists to synthesize real drug carriers for ZOD drug based on calix[$n$]resorcinarenes as well as contribute to opening a new road for designing effective drug delivery systems for ZOD drug.

\section*{\label{auth}Author information}
\textbf{Corresponding Author} \\
*E-mail: ryongnam14@yahoo.com. \\
\textbf{Author Contributions}\\
C.-J. Yu and Y.-M. Jang contributed equally to work presented in this manuscript.\\
\textbf{Notes} \\
The authors declare no competing financial interest.

\section*{\label{ack}Acknowledgments}
This work is supported as part of the fundamental research project ``Design of Innovative Functional Materials for Energy and Environmental Application'' (no. 2016-20) funded by the State Committee of Science and Technology, DPR Korea. Computation was done on the HP Blade System C7000 (HP BL460c) that is owned by Faculty of Materials Science, Kim Il Sung University.

\section*{Abbreviations Used}
ZOD, zoledronate; BP, bisphosphonate; HAP, hydroxyapatite; Cal, calixarene; Res, resorcinarene; P-, phosphonated; S-, sulfonated; -m, methyl; PAO, pseudo-atomic orbital; DZP, double-zeta polarized; GGA, generalized gradient approximation; vdW, van der Waals

\bibliographystyle{achemso}
\bibliography{Reference}

\end{document}